\documentclass[12pt]{article}
\usepackage[latin9]{inputenc}
\usepackage{amsmath}
\usepackage{amssymb}
\usepackage{graphicx}
\usepackage{esint}

\makeatletter

\providecommand{\tabularnewline}{\\}

\usepackage{latexsym}

\graphicspath{{.}}
\makeatother

\begin{document}

\title{Self inductance of a wire loop as a curve integral}

\author{R. Dengler \\
\textit{Rohde}\&\textit{\ Schwarz GmbH}\&\textit{\ Co KG, 1GP1}\\
 \textit{Mühldorfstr. 15, 81671 Munich, P.O.B. 801469.}}
\maketitle
\begin{abstract}
It is shown that the self inductance of a wire loop may be written
as a curve integral akin to the Neumann formula for the mutual inductance
of two wire loops. The only difference is that contributions where
the two integration variables get too close to each other must be
excluded from the curve integral and evaluated in detail. The contributions
of these excluded segments depend on the distribution of the current
in the cross section of the wire. They add to a simple constant proportional
to the wire length. The error of the new expression is of first order
in the wire radius if there are sharp corners and of second order
in the wire radius for smooth wire loops. 
\end{abstract}

\section{Introduction}

\label{sec:Sec_Intro} Electrical inductance plays a crucial role
in power plants, transformers and electronic devices. The coefficients
of self and mutual inductance required to quantitatively describe
inductance belong to the field of magnetostatics. Calculating inductance
coefficients with analytic techniques is, however, impossible except
in simple cases. The mathematical reason for the difficulty is that
the Laplace equation allows analytic solutions only for some symmetric
constellations. There thus are only a few closed-form expressions
for these coefficients. In practice one often is forced to use approximations,
finite element methods or other numerical techniques. The situation
simplifies when the current flows in thin wires. This situation is
analogous to an electrostatic system of point charges, where electric
field and electrostatic energy directly follow from the given charge
distribution, while in a generic system charge and current distributions
also are unknown at the outset.

The purpose of this article is to derive a new expression for the
self inductance of a wire loop, giving self inductance as a curve
integral similar to the Neumann formula for mutual inductance. The
starting point is the expression 
\begin{equation}
W=\frac{\mu_{0}}{8\pi}\int\frac{\mathbf{j}\left(\mathbf{x}\right)\cdot\mathbf{j}\left(\mathbf{x}^{\prime}\right)}{|\mathbf{x-x}^{\prime}|}d^{3}xd^{3}x^{\prime}\label{Eq_W_jj}
\end{equation}
 for the magnetic field energy of a system with current density $\mathbf{j}\left(\mathbf{x}\right)$,
where $\mu_{0}$ is the magnetic constant.\cite{JA75} This expression
essentially was already given by Neumann in 1845.\cite{NEU1847} It
resembles the expression for gravitational or electrostatic potential
energy, the only new ingredient is the scalar product between the
current elements.

For a current density $\mathbf{j}\left(\mathbf{x}\right)=\sum I_{m}\mathbf{j}_{m}\left(\mathbf{x}\right)$
corresponding to $N$ separate current loops with currents $I_{m}$
and \emph{normalized} current densities $\mathbf{j}_{m}$ it follows

\begin{equation}
W=\frac{\mu_{0}}{8\pi}\sum_{m,n=1}^{N}I_{m}I_{n}\int\frac{\mathbf{j}_{m}\left(\mathbf{x}\right)\cdot\mathbf{j}_{n}\left(\mathbf{x}^{\prime}\right)}{|\mathbf{x-x}^{\prime}|}d^{3}xd^{3}x^{\prime}\overset{!}{=}\frac{1}{2}\sum_{m,n=1}^{N}L_{m,n}I_{m}I_{n}.\label{Eq_W}
\end{equation}
If the currents flow in thin wires, then the integrals become curve
integrals, and one immediately reads off the Neumann expression for
mutual inductance of two (filamentary) current loops\cite{NEU1847}
\begin{equation}
L_{1,2}=\frac{\mu_{0}}{4\pi}\oint\frac{d\mathbf{x}_{1}\cdot d\mathbf{x}_{2}}{\left\vert \mathbf{x}_{1}\mathbf{-x}_{2}\right\vert }.\label{Eq_NeumannMutual}
\end{equation}
It is plausible that there exists a similar expression for the self
inductance of a wire loop, but we were not able to find any hint in
the literature. Formally one might read off from equation (\ref{Eq_W})
an expression similar to equation (\ref{Eq_NeumannMutual}), where
the two closed curves coincide. But this cannot be correct, because
$|\mathbf{x-x}^{\prime}|$ now vanishes and the integral isn't defined.
Instead we will prove
\begin{equation}
L=\frac{\mu_{0}}{4\pi}\left(\oint\frac{d\mathbf{x}\cdot d\mathbf{x}^{\prime}}{\left\vert \mathbf{x-x}^{\prime}\right\vert }\right)_{\left\vert s-s^{\prime}\right\vert >a/2}+\frac{\mu_{0}}{4\pi}lY+...\label{Eq_L_Curve}
\end{equation}
where $a$ denotes the wire radius and $l$ the length of the wire.
The variable $s$ measures the length along the wire axis. The constant
$Y$ depends on the distribution of the current in the cross section
of the wire: $Y=0$ if the current flows in the wire surface, $Y=1/2$
when the current is homogeneous across the wire. The ellipses represents
terms like $O\left(\mu_{0}a\right)$ and $O\left(\mu_{0}a^{2}/l\right)$,
which are negligible for $l\gg a$.

In hindsight it is completely natural to use a cutoff of order $a$
in the curve integral. In fact, the exact value of this cutoff is
arbitrary, because the contribution proportional to $lY$ also depends
on this cutoff. The simplest way to determine $Y$ would be to compare
the expression with the self inductance of a long rectangle.

\section{Simple derivation}

\label{sec:Sec_SimpleDerivation} 
\begin{figure}
\centering \includegraphics[width=0.5\textwidth]{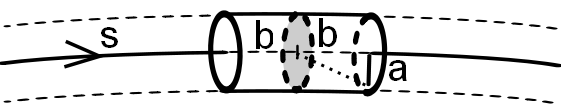} \caption{A section of a wire with radius $a$, with a segment of length $2b$,
and a plane perpendicular to the wire axis at the center of the segment.\label{Fig_WireSegment}}
\end{figure}

Consider equation (\ref{Eq_W})
with $N=1$ for a thin wire with circular cross section, radius $a$
and length $l$. Let $s$ denote the length along the axis of the
wire. The planes perpendicular to the wire axis then define a projection
from the bulk of the wire onto the axis, $\mathbf{x}\rightarrow s\left(\mathbf{x}\right)$
. Selecting a length scale $b$ satisfying $a\ll b\ll l$ allows to
write $L=\left(\mu_{0}/4\pi\right)\left(\overline{L}+\widehat{L}\right)$
with
\begin{eqnarray}
\overline{L} & = & \left(\int\frac{\mathbf{j}\left(\mathbf{x}\right)\mathbf{j}\left(\mathbf{x}^{\prime}\right)}{|\mathbf{x-x}^{\prime}|}d^{3}xd^{3}x^{\prime}\right)_{\left\vert s-s^{\prime}\right\vert >b},\label{Eq_L12_Start}\\
\widehat{L} & = & \left(\int\frac{\mathbf{j}\left(\mathbf{x}\right)\mathbf{j}\left(\mathbf{x}^{\prime}\right)}{|\mathbf{x-x}^{\prime}|}d^{3}xd^{3}x^{\prime}\right)_{\left\vert s-s^{\prime}\right\vert <b}.\notag
\end{eqnarray}
The second part contains contributions from all point pairs $\left\{ \mathbf{x},\mathbf{x}^{\prime}\right\} $
with a distance along the axis smaller than $b$, the first the complement
of this set ($s$ is a cyclic quantity). For given $\mathbf{x}$ the
planes at $s\left(\mathbf{x}\right)\pm b$ delimitate the points $\mathbf{x}^{\prime}$
contributing to the first or second integral, see figure (\ref{Fig_WireSegment}).
$\overline{L}$ now approximately becomes a curve integral and $\widehat{L}$
essentially consists of cylinders of length $2b$.

The strategy then is to replace $\overline{L}$ with the curve integral
and to explicitly evaluate the contribution of the cylinders in $\widehat{L}$.
The cylinders are long in comparison to the radius because of $a\ll b$
and straight (at least most of them) because of $b\ll l$. Actually
the only requirement for the lengths is $a\ll l$, the length $b=\sqrt{al}$
then satisfies $a\ll b\ll l$. The approximation thus is exact in
the limit $a\ll l$ except in special cases. Inserting $\widehat{L}_{0}$
for a straight segment from equation (\ref{Eq_L_Hat_0_Result}) in
the appendix thus leads to
\begin{equation}
L=\frac{\mu_{0}}{4\pi}\left(\oint\frac{d\mathbf{x}\cdot d\mathbf{x}^{\prime}}{|\mathbf{x-x}^{\prime}|}\right)_{\left\vert s-s^{\prime}\right\vert >b}+\frac{\mu_{0}l}{2\pi}\left(\ln\left(\frac{2b}{a}\right)+\frac{Y}{2}\right)+...\label{Eq_L_Result_0}
\end{equation}
This expression cannot depend on the (more or less) arbitrary length
scale $b$. The curve integral thus is%
\footnote{This also can easily be verified with an explicit calculation. %
} $\overline{L}\left(b\right)=const-\frac{\mu_{0}l}{2\pi}\ln\left(2b/b_{0}\right)$.
But $b$ is the only ``short'' length scale in the curve integral,
and $\overline{L}\left(b\right)$ thus also is valid for $b=a/2$.
The expression (\ref{Eq_L_Result_0}) therefore doesn't change if
one formally sets $b=a/2$. Equation (\ref{Eq_L_Result_0}) now agrees
with equation (\ref{Eq_L_Curve}), but some questions remain.

First of all, how accurate is formula (\ref{Eq_L_Curve})? The curve
integral is a purely geometric quantity with dimension ``length''
and order of magnitude $l$. Plausible expressions for the order of
magnitude of the relative error are $a/l$, $\left(a/l\right)^{2}$
and $\left(a/R\right)^{2}$, with $R$ a typical curvature radius
of the wire loop. Errors of this order normally are negligible (and
also occur in the Neumann formula for mutual inductance). But the
derivation of formula (\ref{Eq_L_Curve}) is not as straightforward,
and so what are the actual limits or exceptions?

\section{Examples and comparison with exact self inductance}

\label{sec:Sec_Examples} To get an impression of the accuracy we
have compared self-inductances calculated with the curve integral
with the result of a numeric evaluation of the nominally $6$-dimensional
integral in equation (\ref{Eq_W}). This integral becomes $4$-dimensional
if the currents flow in the wire surface (skin effect, $Y=0$), and
the results below correspond to this skin effect case. The order of
magnitudes of the error terms identified here are corroborated below
in a more detailed derivation of formula (\ref{Eq_L_Curve}).

\subsection*{Straight segment}

The first example is a straight segment with length $c$ and complete
skin effect. This of course isn't a closed circuit, but it might be
an edge of a rectangle. The orthogonal edges of the rectangle don't
interact with the segment because of the scalar product $\mathbf{j}\cdot\mathbf{j}^{\prime}$.
What is missing for a rectangle are the interaction terms with the
opposite edges (and the small contributions from the corners). In
this case the volume integral (\ref{Eq_W}) may even be evaluated
analytically with the result
\[
L=\frac{\mu_{0}}{4\pi}\left\{ 2c\left[\ln\left(\frac{2c}{a}\right)-1\right]+8a/\pi-a^{2}/c+...\right\} ,
\]
while the curve integral (\ref{Eq_L_Curve}) leads to
\begin{equation}
L_{c}\left(c\right)=\frac{\mu_{0}}{4\pi}\left\{ 2c\left[\ln\left(\frac{2c}{a}\right)-1\right]+a\right\} .\label{Eq_LCurve_StraightSegment}
\end{equation}
The difference is of order $O\left(\mu_{0}a\right)$, much smaller
than $\mu_{0}c$ for $c\gg a$.

\subsection*{Circular loop}

The next example is a ring with radius $R$. The curve integral (\ref{Eq_L_Curve})
gives
\[
L_{c}=\mu_{0}R\left(\ln\left(8R/a\right)-2+Y/2\right)+\mu_{0}O\left(a^{2}/R\right).
\]
This expression also may be found in the literature, derived with
the help of elliptic functions and some approximations in a much more
complicated way. The table displays the ratio of the exact inductance
and $L_{c}$ for some values of $R/a$,

\begin{tabular}{l|lllllll}
$R/a$  & $1$  & $2$  & $3$  & $5$  & $10$  & $20$ \tabularnewline
\hline 
$L/L_{c}$  & $5.971$  & $1.238$ & $1.092$  & $1.031$  & $1.00756$ & $1.00188$ \tabularnewline
\end{tabular}\bigskip{}

\noindent The expression $L_{c}$ is more accurate than one might
expect. It gives a reasonable approximation already for $R=3a$, and
the error roughly decays like $O\left(a^{2}/R^{2}\right)$.

\subsection*{Rectangle}

This case is more complicated because in principle also the shape
of the corners comes into play (curvature radius?). But the simplest
thing to do is to evaluate the curve integral (\ref{Eq_L_Curve})
for a rectangle with edges of length $c$ and $d$. Orthogonal edges
decouple because of the scalar product $\mathbf{j}\cdot\mathbf{j}^{\prime}$,
and the first contribution are the terms (\ref{Eq_LCurve_StraightSegment})
for the four edges by themselves. The second contribution are the
parts of the curve integral (\ref{Eq_L_Curve}) with $\mathbf{x}$
on one edge and $\mathbf{x}^{\prime}$ on the opposite one. The condition
$\left\vert s-s^{\prime}\right\vert >a/2$ is irrelevant for these
cross terms, and one easily obtains for parallel edges of length $c$
and distance $d$
\[
L_{c}\left(c,d\right)=\frac{\mu_{0}}{4\pi}\left(4\sqrt{c^{2}+d^{2}}-4d-4c\operatorname{asinh}\left(c/d\right)\right),
\]
and the sum together with the $Y$-term of equation (\ref{Eq_L_Curve})
is
\begin{eqnarray}
L_{c} & = & \frac{\mu_{0}}{\pi}\{\; c\ln\frac{2c}{a}+d\ln\frac{2d}{a}-\left(c+d\right)\left(2-Y/2\right)\label{Eq_L_Rect}\\
 &  & +2\sqrt{c^{2}+d^{2}}-c\operatorname{asinh}\left(c/d\right)-d\operatorname{asinh}\left(d/c\right)+a\;\}.\notag
\end{eqnarray}
This expression also may be found in the literature, with sometimes
a factor $2$ at the $a$-term.\cite{RO1907} The table displays the
ratio of the numerically evaluated self-inductance $L$ of a square
with border length $c$ and corners with curvature radius $a$ and
the curve integral $L_{c}$ for different border length $c$,

\begin{tabular}{l|llllll}
$c/a$  & $5$  & $10$  & $20$  & $40$  & $80$  & $160$ \tabularnewline
\hline 
$L/L_{c}$  & $1.168$  & $1.056$  & $1.0205$ & $1.00776$ & $1.00345$ & $1.00144$\tabularnewline
$\left(L-L_{c}\right)/\mu_{0}a$  & $0.501$  & $0.562$ & $0.5871$ & $0.57826$ & $0.63559$ & $0.63185$\tabularnewline
$L/L_{c}^{\text{exact}}$  & $1.053$ & $1.021$ & $1.0080$ & $1.0029$ & $1.00147$ & $1.00061$\tabularnewline
\end{tabular}\bigskip{}

\noindent The curvature radius $a$ is minimal in that the centre
of curvature lies on the inner border of the wire. It is remarkable
that the absolute error nearly remains constant. The last row of the
table displays the ratio of the exact self inductance and the exact
curve integral (\ref{Eq_L_Curve}) (with round corners), also evaluated
numerically. This expression is a better approximation for small $c/a$,
where the square with round corners degenerates to a ring.

\subsection*{Equilateral triangle}

The curve integral (\ref{Eq_L_Result_0}) for an equilateral triangle
with edge length $c$ consists of three times the expression (\ref{Eq_L_Curve})
for the edges by themselves and three times the interaction energy
$L_{c}\left(c,c,120\right)$ of adjacent edges (with $s$ on one edge
and $s^{\prime}$ on the other, see appendix \ref{sec:App_Adjacent}.
There is no such interaction for rectangles because of the scalar
product),
\[
L_{c}=\frac{\mu_{0}}{2\pi}3c\left\{ \ln\left(\frac{c}{a}\right)-1-\ln\frac{3}{2}\right\} .
\]
The table displays the ratio of the exact self-inductance $L$ of
an equilateral triangle with border length $c$ and corners with curvature
radius $a$ and the curve integral $L_{c}$ for different border length
$c$,

\begin{tabular}{l|llllll}
$c/a$  & $5$  & $10$  & $20$  & $40$  & $80$  & $160$ \tabularnewline
\hline 
$L/L_{c}$  & $2.983$  & $1.2534$ & $1.0743$ & $1.0264$ & $1.0102$ & $1.0043$\tabularnewline
$\left(L-L_{c}\right)/\mu_{0}a$  & $0.966$ & $1.0856$ & $1.1288$ & $1.1501$ & $1.1599$ & $1.2017$\tabularnewline
$L/L_{c}^{\text{exact}}$  & $1.376$ & $1.0931$ & $1.0302$ & $1.0110$ & $1.0043$ & $1.0019$\tabularnewline
\end{tabular}\bigskip{}

\noindent The absolute error $\left(L-L_{c}\right)/\mu_{0}$ is nearly
constant also here. The last row again is the ratio of the exact self
inductance and the curve integral (\ref{Eq_L_Curve}) with round corners,
evaluated numerically.

\subsection*{Parallel wires}

For a loop consisting of infinitely long parallel wires the condition
$a\ll l$ is perfectly met and the curve integral gives the exact
self inductance even for minimal distance $d=2a,$ 
\[
L_{c}=\frac{\mu_{0}l}{\pi}\left(\ln\frac{d}{a}+Y/2\right).
\]
This expression is the limiting case of the expression (\ref{Eq_L_Rect})
for a long rectangle. The point is, that the contribution of the corners
becomes negligible for a long rectangle, and that the replacement
of the (circular symmetric) current by a line current doesn't change
the magnetic field according to Ampere's law. Of course, the assumption
of a circular symmetric current distribution gets wrong in the skin
effect case if the wires are close to each other because of additional
screening currents.

To summarize, formula (\ref{Eq_L_Result_0}) is rather accurate even
for circuits with a linear extension as small as $20$ times the wire
radius, even if the circuit contains sharp corners.

\section{Error estimation}

\label{sec:Sec_ErrorEstimation} 
\begin{figure}
\centering \includegraphics[width=0.5\textwidth]{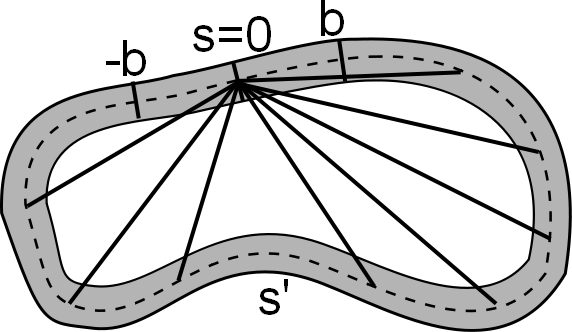} \caption{A wire loop and distances relevant for calculating the energy of a
disc at $s=0$ relative to other points in the wire.\label{Fig_Loop_s_}}
\end{figure}

According to equation (\ref{Eq_L12_Start}) the self inductance may
be written as $L=\left(\mu_{0}/4\pi\right)\left(\overline{L}+\widehat{L}\right)$,
where $\widehat{L}$ contains the short segments and $\overline{L}$
the complement. The self inductance $L$ of course doesn't depend
on the arbitrary segment length $b$.

Let us now introduce some notation. We use a coordinate system $\left\{ s,r,\phi\right\} $
in the wire where the length $s$ along the wire axis is cyclic with
period $l$, and the coordinates $\left\{ r,\phi\right\} $ describe
planes perpendicular to the wire axis. The intersections of the planes
and the wire are assumed to be circular, $0\leqq r\leqq a$ and $0\leqq\phi\leqq2\pi$.
The volume element reads $dV=\left(1+r\cos\phi/R\left(s\right)\right)rdrd\phi ds$,
where $R\left(s\right)$ denotes the curvature radius of the wire,
and $\phi=0$ at the outer border of the wire (this is possible at
least locally). The volume element also may be written as $dV=dsd\widetilde{A}$
with $d\widetilde{A}=\left(1+r\cos\phi/R\left(s\right)\right)dA$
and area element $dA=rdrd\phi$. The coordinates become cylindrical
coordinates for straight wire segments, i.e. for $R=\infty$.

The current density $\mathbf{j}$ is normalized, that is $\int dA\left\vert \mathbf{j}\right\vert \mathbf{=}\int d\widetilde{A}\left\vert \mathbf{j}\right\vert =1$.
We will also need the radial moments
\[
a_{n}=\left\langle r^{n}\right\rangle =\int d\widetilde{A}r^{n}\left\vert \mathbf{j}\right\vert 
\]
of the current distribution. In the skin effect case of course $a_{n}=a^{n}$.

One quantity of interest then is
\begin{equation}
\overline{L}\left(s\right)=\int ds^{\prime}\theta\left(\left\vert s^{\prime}-s\right\vert -b\right)\int d\widetilde{A}d\widetilde{A}^{\prime}\frac{\mathbf{j}\left(s,r,\phi\right)\cdot\mathbf{j}\left(\mathbf{x}^{\prime}\right)}{|\mathbf{x}\left(s,r,\phi\right)\mathbf{-x}\left(s^{\prime},r^{\prime},\phi^{\prime}\right)|},\label{Eq_LBar_0}
\end{equation}
the energy of the current in the plane at $s=0$ with respect to the
current at $\left\vert s^{\prime}-s\right\vert >b$, see figure (\ref{Fig_Loop_s_}).
The symbol $\theta$ denotes the step function. To obtain $\overline{L}$
from $\overline{L}\left(s\right)$ requires to integrate over $s$.
The curve integral
\begin{equation}
\overline{L}_{\gamma}\left(s\right)=\oint ds^{\prime}\theta\left(\left\vert s^{\prime}-s\right\vert -b\right)\frac{\cos\left(\mathbf{j}\left(s\right),\mathbf{j}\left(s^{\prime}\right)\right)}{|\mathbf{x}\left(s,0,0\right)\mathbf{-x}\left(s^{\prime},0,0\right)|},\label{Eq_LBar_Gamma}
\end{equation}
is an approximation for $\overline{L}\left(s\right)$.

Similarly we write the short segment around the plane at $s=0$ as
\begin{equation}
\widehat{L}\left(s\right)=\int ds^{\prime}\theta\left(b-\left\vert s^{\prime}-s\right\vert \right)\int d\widetilde{A}d\widetilde{A}^{\prime}\frac{\mathbf{j}\left(s,r,\phi\right)\cdot\mathbf{j}\left(s^{\prime},r^{\prime},\phi^{\prime}\right)}{|\mathbf{x}\left(s,r,\phi\right)\mathbf{-x}\left(s^{\prime},r^{\prime},\phi^{\prime}\right)|},\label{Eq_L_Hat_0}
\end{equation}
These definitions allow to write
\begin{eqnarray}
\frac{4\pi}{\mu_{0}}L\left(s\right) & = & \overline{L}\left(s\right)+\widehat{L}\left(s\right)=\label{Eq_L_0}\\
 & = & \left\{ \left(\overline{L}_{\gamma}+\widehat{L}_{\gamma}\right)+\left(\overline{L}-\overline{L}_{\gamma}+\hat{P}_{0}\right)+\left(\widehat{L}-\widehat{L}_{0}\right)\right\} _{s},\notag
\end{eqnarray}
where we have added and subtracted the curve integral $\overline{L}_{\gamma}\left(s\right)$,
the segment integral $\widehat{L}_{0}\left(s\right)$ for a straight
segment from equation (\ref{Eq_L_Hat_0_Result}) and the approximation
$\widehat{L}_{\gamma}\left(s\right)=2\left(\ln\left(2b/a\right)+Y/2\right)$
for $\widehat{L}_{0}\left(s\right)$. The first bracket in equation
(\ref{Eq_L_0}) now is formula (\ref{Eq_L_Result_0}), and the two
other brackets thus represent the error. The second bracket contains
the difference of the volume and the curve integral (for $\left\vert s^{\prime}-s\right\vert >b$)
plus the power series $\hat{P}_{0}=\widehat{L}_{0}-\widehat{L}_{\gamma}$
from equation (\ref{Eq_L_Hat_0_Result}), the third bracket is the
difference of the actual segment integral (\ref{Eq_L_Hat_0}) and
the segment integral for a straight segment. It is evident that the
error becomes small in suitable limits, and we now want to determine
the order of magnitude of the error.

\subsection*{Smooth current loops}


We first consider smooth current loops, that is current loops with
a minimal curvature radius $R$ comparable with the system size. We
also assume that the loop returns immediately and doesn't touch itself
anywhere in between. Such complications are considered below.

The coordinates $\mathbf{x}$ and $\mathbf{x}^{\prime}$ in the integral
(\ref{Eq_LBar_0}) above may be expanded like $\mathbf{x}=\mathbf{x}\left(s,0,0\right)+\mathbf{x}_{1}\left(s,r,\phi\right)$,
and thus
\begin{eqnarray*}
\mathbf{x}-\mathbf{x}^{\prime} & = & \mathbf{x}_{s,s^{\prime}}+\mathbf{x}_{1}-\mathbf{x}_{1}^{\prime},\\
\left(\mathbf{x}-\mathbf{x}^{\prime}\right)^{2} & = & x_{s,s^{\prime}}^{2}+2\mu x_{s,s^{\prime}}+\nu^{2}
\end{eqnarray*}
with $\mathbf{x}_{s,s^{\prime}}=\mathbf{x}\left(s,0,0\right)-\mathbf{x}\left(s^{\prime},0,0\right)$
the distance of the projections onto the axis and $\left\vert \mathbf{x}_{1}\right\vert $\ and
$\left\vert \mathbf{x}_{1}^{\prime}\right\vert $ of order $O\left(a\right)$.
The abbreviations are
\begin{eqnarray*}
\mu & = & \widehat{\mathbf{x}}_{ss^{\prime}}\cdot\mathbf{x}_{1}+\widehat{\mathbf{x}}_{s^{\prime}s}\cdot\mathbf{x}_{1}^{\prime},\\
\nu^{2} & = & \left(\mathbf{x}_{1}-\mathbf{x}_{1}^{\prime}\right)^{2}.
\end{eqnarray*}
The procedure now is to use the multipole expansion
\begin{eqnarray}
\frac{1}{\left\vert \mathbf{x}-\mathbf{x}^{\prime}\right\vert } & = & \frac{1}{x_{s,s^{\prime}}}-\frac{\mu}{x_{s,s^{\prime}}^{2}}+\frac{1}{2x_{s,s^{\prime}}^{3}}\left(3\mu^{2}-\nu^{2}\right)-\frac{1}{2x_{s,s^{\prime}}^{4}}\left(5\mu^{3}-3\mu\nu^{2}\right)\label{Eq_MultipoleExp}\\
 &  & +\frac{1}{8x_{s,s^{\prime}}^{5}}\left(35\mu^{4}-30\mu^{2}\nu^{2}+3\nu^{4}\right)+O\left(a^{5}x_{s,s^{\prime}}^{-6}\right).\notag
\end{eqnarray}
This expansion converges for $x_{s,s^{\prime}}\geqq\left\vert \mathbf{x}_{1}^{\prime}\mathbf{-x}_{1}\right\vert $,
the coefficients are the coefficients of the Legendre polynomials.
Inserting the leading \emph{monopole} term $1/x_{s,s^{\prime}}$ of
equation (\ref{Eq_MultipoleExp}) into equation (\ref{Eq_LBar_0})
reproduces the curve integral $\overline{L}_{\gamma}\left(s\right)$.
The higher multipole terms describe the difference between the volume
integral $\overline{L}\left(s\right)$ and the curve integral. The
$s^{\prime}$-integral in the multipole terms converges and the difference
thus is mainly a local quantity. The nominal order of magnitude of
the multipole terms $\overline{L}_{n}$ is $a^{n}/b^{n}$, $n\geqq1$.
With a length $b=\sqrt{aR}$ the order of magnitude becomes $\left(a/R\right)^{n/2}$,
and the expansion up to the hexadecupole ($n=4$) is needed to get
the $\left(a/R\right)^{2}$ approximation.

The volume element is
\[
\left(1+r\cos\phi/R\left(s\right)\right)rdrd\phi ds=\left(1+\mathbf{x}_{1}\cdot\mathbf{R}\left(s\right)/R^{2}\left(s\right)\right)rdrd\phi ds,
\]
where $\mathbf{R}\left(s\right)$ is the local curvature radius vector.
Integrals over $\phi$ and $\phi^{\prime}$ then may be evaluated
with the help of
\begin{eqnarray*}
\left\langle \mathbf{x}_{1}\right\rangle  & = & 0,\\
\left\langle \left(\mathbf{x}_{1}\right)_{m}\left(\mathbf{x}_{1}\right)_{n}\right\rangle  & = & r^{2}P_{m,n}/2,
\end{eqnarray*}
where $P_{m,n}$ is the projection operator projecting onto the plane
perpendicular to the wire axis. This implies $\mathbf{P\cdot R}=\mathbf{R}$.
Inserting now the multipole expansion (\ref{Eq_MultipoleExp}) into
the volume integral (\ref{Eq_LBar_0}) generates an expansion of the
difference of the volume and the curve integral in the region $\left\vert s^{\prime}\right\vert >b$.

\paragraph{Dipole}

The voltage drops by the same amount along inner and outer border
along curved parts of the loop. Electric field and current density
thus are larger at the inner border, and the current density thus
depends on $r$ and $\varphi$. This $\varphi$-dependence of the
current density compensates the $\varphi$-dependence $1+r\cos\left(\varphi\right)/R$
from the volume element. The simple result then is: there is no dipole
contribution. The factors $\cos\varphi$ and $\cos\varphi'$ from
$\mu$ give $0$ after integration over the angles. There are in fact
short transition regions between straight and curved parts of the
loop where the current distribution changes from uniform to non-uniform,
but this only leads to corrections of higher order.

\paragraph{Quadrupole}

After integration over the angles $\phi$ and the radial coordinates
$r$ there remains
\[
\overline{L}_{2}\left(s\right)=a_{2}\oint ds^{\prime}\theta\left(\left\vert s^{\prime}-s\right\vert -b\right)\left(\frac{3}{4}\left(2-\left(\widehat{\mathbf{n}}\cdot\widehat{\mathbf{x}}_{s,s^{\prime}}\right)^{2}-\left(\widehat{\mathbf{n}}^{\prime}\cdot\widehat{\mathbf{x}}_{s,s^{\prime}}\right)^{2}\right)-1\right)\frac{\cos\alpha}{x_{s,s^{\prime}}^{3}},
\]
where $\widehat{\mathbf{n}}$ denotes a unit vector in the direction
of the wire axis and the expression $\mathbf{P}=1-\widehat{\mathbf{n}}\widehat{\mathbf{n}}$
for the projection operator was used. In $\overline{L}_{2}\left(s\right)$
one may recognize $1-\left(\widehat{\mathbf{n}}\cdot\widehat{\mathbf{x}}_{s,s^{\prime}}\right)^{2}=\sin^{2}\psi$
and $1-\left(\widehat{\mathbf{n}}^{\prime}\cdot\widehat{\mathbf{x}}_{s,s^{\prime}}\right)^{2}=\sin^{2}\psi^{\prime}$,
where $\psi$ denotes the angle between the distance vector $\mathbf{x}_{s,s^{\prime}}$
and the wire axis. With $\psi^{\prime}=\psi=2\alpha=\left(s-s\right)^{\prime}/2R$
for a smooth current loop there remains an error $\left(a^{2}/R^{2}\right)\ln\left(R/b\right)$.
The $-a_{2}/b^{2}$ term (from the $-1$) is peculiar. With the choice
$b=\sqrt{aR}$ it would be of order $a/R$, but it gets cancelled
against the leading term of the power series $\hat{P}_{0}$ of equation
(\ref{Eq_L_0}). This of course is the reason for combining $\hat{P}_{0}$
with the multipole expansion in equation (\ref{Eq_L_0}): for a straight
segment the formula is exact, and the third bracket in equation (\ref{Eq_L_0})
vanishes. The multipole expansion together with $\hat{P}_{0}$ thus
also vanishes for $R=\infty$.

\paragraph{Oktupole}

The oktupole contributes at most a term of order $a^{3}/b^{3}$. \ But
the $\cos\phi$ and $\sin\psi$ from the odd power of $\mu$ make
the actual contribution smaller than the expected $\sim a^{2}/R^{2}$.

\paragraph{Hexadecupole}

The hexadecupole term is of order $a^{4}/b^{4}\sim a^{2}/R^{2}$ and
its leading part must be kept. The $\mu$ factors contain a factor
$\sin\psi\sim b/R$ and may be dropped. The $\nu^{4}$ leads to
\[
\overline{L}_{4}\left(s\right)=\frac{3}{8}\oint ds^{\prime}\theta\left(\left\vert s^{\prime}-s\right\vert -b\right)\left[2a_{4}+a_{2}^{2}\left(3+\cos^{2}\alpha\right)\right]\frac{\cos\alpha}{x_{s,s^{\prime}}^{5}}.
\]
The factors $\cos\alpha$ may be replaced with $1$ because of $\alpha=\left(s^{\prime}-s\right)/R$
is small. The integral converges and contributes an error like $a^{2}/R^{2}$.
But this term gets cancelled by the second term of the power series
$\hat{P}_{0}$ from equation (\ref{Eq_L_Hat_0_Result}).

For smooth current loops it finally follows
\begin{equation}
\frac{4\pi}{\mu_{0}}L\left(s\right)=\left(\overline{L}_{\gamma}+\widehat{L}_{\gamma}\right)+O\left(\left(a^{2}/R\right)\ln\frac{R}{b}\right)+\left(\widehat{L}-\widehat{L}_{0}\right),\label{Eq_L_Smooth_Final}
\end{equation}
where the first bracket on the r.h.s is formula (\ref{Eq_L_Result_0}),
evaluated with $b=\sqrt{aR}$. The remaining segment integral $\widehat{L}$
for a segment with curvature radius $R$ is evaluated in appendix
\ref{sec:App_Short_Curved}. The result (\ref{Eq_L_Hat_R_0}) contributes
another logarithmic error $\left(a^{2}/R^{2}\right)\ln\left(R/b\right)$,
and a larger error of order $b^{2}/R^{2}\thicksim a/R$. We now drop
all errors of order $O\left(\left(a^{2}/R\right)\ln\left(R/b\right)\right)$
and $O\left(a^{2}/R\right)$ from the multipole expansion and the
curved segment . 

As in the simple derivation of the formula above one may notice that
there is no small lenght scale in the formula $\bar{L}_{\gamma}+\widehat{L}_{\gamma}$
and its $b$-dependence is under control for all $b<\sqrt{aR}$ and
given by (\ref{Eq_LBar_Gamma_R}). It contains a ``large'' $b^{2}/R^{2}\thicksim a/R$
term. The essential point now is that this $b$-dependence of the
formula exactly absorbs the remaining large error from $\widehat{L}$
(the cancellation comes about, because $b$ is the lower limit in
the curve integral and the upper limit in the volume integral). Nothing
therefore changes if we now set $b=a/2$, except that there only remains
formula (\ref{Eq_L_Result_0}) evaluated for $b=a/2$ and small terms
of order $O\left(\left(a^{2}/R\right)\ln\left(R/a\right)\right)$
and $O\left(a^{2}/R\right)$. 

For a circular loop the leading error terms can even be evaluated
in closed form. The results for the self-inductances in the skin effect
and volume current case are
\begin{eqnarray*}
L_{S} & = & \mu_{0}R\left\{ \left(1+\frac{3}{4}\frac{a^{2}}{R^{2}}\right)\ln\left(\frac{8R}{a}\right)-2-\frac{3}{2}\frac{a^{2}}{R^{2}}\right\} ,\\
L_{V} & = & \mu_{0}R\left\{ \left(1+\frac{3}{8}\frac{a^{2}}{R^{2}}\right)\ln\left(\frac{8R}{a}\right)-1.75-\frac{2}{3}\frac{a^{2}}{R^{2}}\right\} .
\end{eqnarray*}
The corrections perfectly agree with a numeric evaluation of the $6$-
or $4$-dimensional integrals, but strongly disagree with expressions
in the literature.\cite{GR1948} The reason appears to be that these
calculations assume a constant current distribution in the surface
or cross section of the wire, wrong just in the curved parts of the
loop.

\subsection*{Current loops with sharp corners}

The errors originate from the curved parts of the current loop, and
for current loops with sharp corners one may expect larger errors.
A simple way to construct such loops is to insert straight segments
into a circular loop. It was shown above that the absolute error for
a circular loop with radius $R$ is of order $O\left(\mu_{0}a{}^{2}/R\right)$.
This estimation is valid even for minimal curvature radius $R=a$
(the inductance has dimension $length\times\mu_{0}$ and $a$ is the
only available length for such loops). Since the loop with inserted
straight segments is better approximated by the curve integral the
absolute error can only be smaller. The formal reason is that a filamentary
straight segment generates the same magnetic field as the actual axially
symmetric current distribution.

The inductance of a loop of extension $l\gg a$ generally is of order
$O\left(\mu_{0}l\cdot\ln\left(l/a\right)\right)$. The ratio with
the absolute error leads to a generic estimation of the relative error
of formula (\ref{Eq_L_Curve})
\[
\Delta L/L=\sum_{n}O\left(\frac{a^{2}}{lR_{n}}\right),
\]
where the index $n$ enumerates the corners and the logarithmic factor
is neglected. We have allowed here corners with different curvature
radius $R_{n}$. The special case with curvature radii of order $O\left(l\right)$
leads back to the estimation of the error for smooth current loops
above. Sharp corners with curvature radius $R=O\left(a\right)$ contribute
a relative error of order $O\left(a/l\right)$. Corners with a small
angle of course should get a smaller weight in the sum. This estimation
could be made more rigorous, but the procedure only is circumstantial
and of no interest here.

\subsection*{Current loops not closing immediately}

The error estimations above fail if the current loop comes close to
itself before it actually closes, that is for tight spirals, coils
or something like that (the estimation $x_{s,s^{\prime}}\thicksim\left|s^{\prime}-s\right|$
gets invalid). In this case the replacement of the actual current
with a filamentary current generates additional errors at these positions.
But this isn't specific for equation (\ref{Eq_L_Curve}), exactly
the same (small) errors are contained in the Neumann formula for mutual
inductance. 

The error may easily be estimated, because the condition $\left|s^{\prime}-s\right|>b$
is irrelevant if $s'$ is on one winding and $s$ on another. A simple
possibility is to consider as a worst case scenario two current loops
of radius $R$ a distance $d>2a$ on top of each other. The error
comes from the multipole expansion (\ref{Eq_MultipoleExp}) valid
everywhere now, and it is a simple matter to estimate the integrals.
The order of magnitude of the relative error is $O\left(a^{2}/Rd\right)$,
small except for small curvature radius \textit{and} small distance
(both of order $O\left(a\right)$).

\section{Conclusions}

\label{sec:Sec_Conclusions} The curve integral (\ref{Eq_L_Curve})
for the self inductance of a wire loop is only a little bit more complicated
than the Neumann formula for the mutual inductance of two wire loops.
The exact expression for self inductance is a $6$-dimensional integral
with a logarithmic divergence and several length scales. Nevertheless
clear statements follow for the accuracy of formula (\ref{Eq_L_Curve}),
for loops consisting of straight segments as well as for smooth loops.
The error originates from the curved parts of the loop, and is of
order $\mu_{0}a$ or $\mu_{0}a^{2}/l$, negligible for most practical
purposes. The leading error presumably even might be given as a sum
over the corners for loops consisting of straight segments or as additional
curve integrals for smooth current loops (see the quadrupole contribution
above). The techniques used for error estimation also may be used
for the Neumann formula.

The self inductance curve integral can be evaluated analytically in
many cases, for instance for current loops consisting of coplanar
straight segments (not a new result; see also appendix \ref{sec:App_Adjacent}
and \ref{sec:App_NonAdjacent}). But equation (\ref{Eq_L_Curve})
is valid for abitrary curves, and the numerical evaluation of two-dimensionals
integrals with a computer program is a breeze with appropriate numerical
libraries. The information for the self inductance is contained in
the curve spanned by the current loop, and any self inductance calculation
at least requires a double integral along the curve. Simpler methods
or estimations based on ``partial inductance'' or only the magnetic
flux miss this point and only may work in special cases.

The fact that two coinciding points cause problems in self inductance
calculations is well known and has been circumvented in several ways,
for instance by distributing the current onto two filamentary loops.
But no systematic approximations can be obtained in this way.

Current distributions which are not circular symmetric also lead to
formula (\ref{Eq_L_Curve}), with a cutoff and a constant $Y$ depending
on the current distribution. An example are circuits consisting of
coplaner flat strips of width $w$. The self inductance of such circuits
is given in the accuracy described above by formula (\ref{Eq_L_Curve})
with $a=w$ and $Y=3$.

\appendix
\numberwithin{equation}{section}

\section{Contribution from straight segments of length $2b$}

\label{sec:App_Short_Straight} The contribution $\widehat{L}$ to
the self inductance in equation (\ref{Eq_L12_Start}) is due to the
interaction of the current in the plane $s$ with the current in all
planes $s^{\prime}$ with $\left\vert s^{\prime}-s\right\vert <b$.
This value depends on the current distribution in the wire and on
the wire geometry within the segment $\left[s-b\text{,}s+b\right]$,
but may be evaluated if the segment is straight or slightly curved.
This $s$-dependent value of course still is to be integrated over
all $s$.

To get an approximation for $\widehat{L}$ in the straight wire case
use cylinder coordinates with a length $s$ along the axis and area
element $dA=rdrd\phi$ (see figure (\ref{Fig_WireSegment})). This
leads to
\begin{eqnarray}
\widehat{L}_{0} & = & \oint ds\widehat{L}\left(s\right),\label{Eq_L2_L2bs}\\
\widehat{L}_{0}\left(s\right) & = & \left(\int\frac{\mathbf{j}\left(r\right)\mathbf{j}\left(r^{\prime}\right)}{|\mathbf{x}\left(s,r,\phi\right)\mathbf{-x}^{\prime}|}ds^{\prime}dA^{\prime}dA\right)_{\left\vert s\left(\mathbf{x}^{\prime}\right)-s\right\vert <b}.\notag
\end{eqnarray}
In the latter integral $\mathbf{x}$ extends over the plane through
the centre of a cylindrical segment, $\mathbf{x}^{\prime}$ extends
over the complete segment. The integral $\widehat{L}_{0}\left(s\right)$
of course is independent of $s$. The integral over $s^{\prime}$
(from $-b$ to $b$) may be performed using $|\mathbf{x-x}^{\prime}|^{2}=N^{2}+s^{\prime2}$,
$N^{2}=r^{2}+r^{\prime2}-2rr^{\prime}\cos\left(\phi-\phi^{\prime}\right)$,
\begin{eqnarray*}
\widehat{L}_{0}\left(0\right) & \cong & 2\int dAdA^{\prime}\mathbf{j}\left(r\right)\mathbf{j}\left(r^{\prime}\right)\operatorname{asinh}\left(b/N\right)\\
 & = & 2\int dAdA^{\prime}\mathbf{j}\left(r\right)\mathbf{j}\left(r^{\prime}\right)\left\{ \ln\left(2b/a\right)-\ln\left(N/a\right)+A_{1}\left(N/b\right)+...\right\} 
\end{eqnarray*}
In the second line the expansion
\begin{eqnarray}
\operatorname{asinh}\left(x\right) & = & \ln\left(2x\right)+A_{1}\left(1/x\right)\label{Eq_Asinh_Expansion}\\
A_{1}\left(x\right) & = & \sum\nolimits _{n=1}^{\infty}\frac{1\cdot3...\left(2n-1\right)}{2\cdot4...2n}\frac{\left(-1\right)^{n+1}}{2n}x^{2n}=x^{2}/4-3x^{4}/32+...\notag
\end{eqnarray}
was used. The expansion converges because of $N=O\left(a\right)\ll b$.
It doesn't matter which $\phi^{\prime}$ occurs in the $\phi$-integral
and thus we now set $\phi^{\prime}=0$. Because of $\int dA\left\vert \mathbf{j}\right\vert =1$
the leading term simply becomes $2\ln\left(2b/a\right)$. The second
term follows from $\ln\left(N/a\right)=\frac{1}{2}\ln\left(\rho^{2}+\rho^{\prime2}-2\rho\rho^{\prime}\cos\phi\right)$
with $\rho=r/a$ and $\rho^{\prime}=r^{\prime}/a$ and 
\begin{equation}
\frac{1}{2\pi}\int_{0}^{2\pi}\ln\left(\rho^{2}+\rho^{\prime2}-2\rho\rho^{\prime}\cos\phi\right)d\phi=2\ln\left(\rho_{>}\right),\label{Eq_Bronstein}
\end{equation}
where $\ln\left(\rho_{>}\right)=\theta\left(\rho-\rho^{\prime}\right)\ln\rho+\theta\left(\rho^{\prime}-\rho\right)\ln\rho^{\prime}$.
This term thus vanishes in the skin effect case where the current
differs from $0$ only for $\rho=\rho'=1$. The current density in
the constant current case is $j\left(r\right)=1/\left(\pi a^{2}\right)$
and the second term becomes
\[
-\frac{\left(2\pi\right)^{2}}{\pi^{2}}2\int_{0}^{1}d\rho\rho\int_{0}^{1}d\rho^{\prime}\rho^{\prime}\ln\left(\rho_{>}\right)=1/2.
\]
A rapidly convergent expansion approximation for $b\gg a$ thus is
\begin{eqnarray}
\widehat{L}_{0}\left(0\right) & = & \widehat{L}_{\gamma}\left(0\right)+\widehat{P}_{0}\left(0\right),\label{Eq_L_Hat_0_Result}\\
\widehat{L}_{\gamma}\left(0\right) & = & 2\ln\left(2b/a\right)+Y,\notag\\
\widehat{P}_{0}\left(0\right) & =\left\langle 2A_{1}\left(\frac{N}{b}\right)\right\rangle = & \frac{a_{2}}{b^{2}}-\frac{3}{8b^{4}}\left(a_{4}+2a_{2}^{2}\right)+O\left(\frac{a^{6}}{b^{6}}\right)\notag
\end{eqnarray}
with $Y=1/2$ for a constant current distribution and $Y=0$ in the
skin effect case.

\section{Contribution from a curved segment}

\label{sec:App_Short_Curved} The goal is to evaluate the integral
$\widehat{L}\left(0\right)$ from equation (\ref{Eq_L_Hat_0}) for
a segment of length $2b$ and constant curvature radius $R$,
\[
\widehat{L}_{R}\left(0\right)=\int\frac{ds^{\prime}d\widetilde{A}d\widetilde{A}^{\prime}}{|\mathbf{x}\left(0,r,\phi\right)\mathbf{-x}^{\prime}|}\theta\left(b-|s^{\prime}|\right)j\left(r\right)j\left(r^{\prime}\right)\cos\left(\frac{s^{\prime}}{R}\right).
\]
The distance up to order $O\left(R^{-2}\right)$ follows from
\begin{eqnarray*}
\left(\mathbf{x-x}^{\prime}\right)^{2} & \cong & s^{\prime2}+N^{2}+q,\\
N^{2} & = & r^{2}+r^{\prime2}-2rr^{\prime}\cos\left(\phi-\phi^{\prime}\right),\\
q & = & s^{\prime2}\left[\left(r^{\prime}\cos\phi^{\prime}+r\cos\phi\right)R^{-1}+rr^{\prime}R^{-2}\cos\phi\cos\phi^{\prime}\right]-s^{\prime4}/\left(12R^{2}\right)+s^{\prime6}/\left(360R^{4}\right).
\end{eqnarray*}
Expanding in $q$ gives
\[
\frac{1}{|\mathbf{x-x}^{\prime}|}=\frac{1}{\left(s^{\prime2}+N^{2}\right)^{1/2}}-\frac{q}{2\left(s^{\prime2}+N^{2}\right)^{3/2}}+\frac{3q^{2}}{8\left(s^{\prime2}+N^{2}\right)^{5/2}}+...
\]
The $N$ in the denominators is negligible for $b\gg a$ and the elementary
integrals lead to
\begin{equation}
\widehat{L}_{R}\left(0\right)-\widehat{L}_{0}\left(0\right)=-\frac{11}{24}\frac{b^{2}}{R^{2}}+O\left(\frac{a_{2}}{R^{2}}\ln\frac{b}{a}\right).\label{Eq_L_Hat_R_0}
\end{equation}
The term of order $O\left(a_{2}/R^{2}\right)$ contributes to the
error of the self inductance formula with the expected order of magnitude.
It is essential however, that the $O\left(b^{2}/R^{2}\right)$ term,
which is of order $a/R$ because of $b=\sqrt{aR}$, cancels against
a contribution from the curve integral $\overline{L}_{\gamma}\left(0\right)$
from equation (\ref{Eq_LBar_Gamma}).

To verify this cancellation start with
\[
\frac{d}{db}\overline{L}_{\gamma}\left(0\right)=\frac{-\cos\left(\mathbf{j}\left(0\right)\mathbf{,j}\left(b\right)\right)}{|\mathbf{x}\left(0\right)\mathbf{-x}\left(b\right)|}-\frac{-\cos\left(\mathbf{j}\left(0\right)\mathbf{,j}\left(-b\right)\right)}{|\mathbf{x}\left(0\right)\mathbf{-x}\left(-b\right)|}.
\]
Inserting $\left\vert x-x^{\prime}\right\vert =2R\sin b/2R$ and $\cos\left(\mathbf{j}\left(0\right)\mathbf{,j}\left(b\right)\right)=\cos b/r$
for a section with curvature radius $R$ gives
\[
\frac{d}{db}\overline{L}_{\gamma}\left(0\right)=\frac{-2}{b}\left(1-\frac{11}{24}\left(\frac{b}{R}\right)^{2}+...\right),
\]
with integral
\begin{equation}
\overline{L}_{\gamma}\left(0\right)=const-2\ln\frac{2b}{a}+\frac{11}{24}\frac{b^{2}}{R^{2}}+...\label{Eq_LBar_Gamma_R}
\end{equation}

\section{Curve integral for adjacent straight segments}

\label{sec:App_Adjacent} For completeness we display here the curve
integral contribution to the self inductance from adjacent straight
segments of length $c$ and $d$ with an angle $\alpha$ between the
currents,
\begin{eqnarray*}
L_{\gamma}\left(c,d,\alpha\right) & = & \frac{\mu_{0}}{2\pi}\cos\left(\alpha\right)\{c\operatorname{asinh}\frac{d+c\cos\alpha}{c\sin\alpha}+d\operatorname{asinh}\frac{c+d\cos\alpha}{d\sin\alpha}\\
 &  & -\left(c+d\right)\operatorname{asinh}\frac{\cos\alpha}{\sin\alpha}-\frac{2b}{\sqrt{1\left(1-\cos\alpha\right)}}\operatorname{asinh}\frac{1-\cos\alpha}{\sin\alpha}\}.
\end{eqnarray*}
For each corner such a term is to be added to the contribution (\ref{Eq_LCurve_StraightSegment})
of the (straight) segments by themselves. The $b$-term is of order
$O\left(a\right)$ for $b=a/2$ and normally may be neglected.

\section{Curve integral for non-adjacent coplanar straight segments}

\label{sec:App_NonAdjacent} 
\begin{figure}
\centering \includegraphics[width=0.5\textwidth]{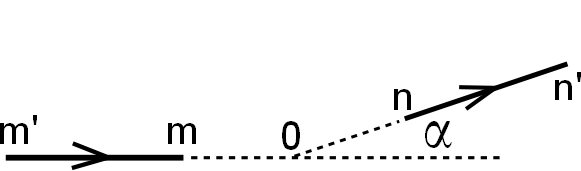} \caption{Two non-adjacent coplanar straight segments. The coordinates $m$,
$m^{\prime}$ and $n$, $n^{\prime}$measure the distance of the end
points from the intersection of the segment extensions.\label{Fig_NonAdjacent}}
\end{figure}

What then is missing for calculating the self inductance of a loop
consisting of arbitrary coplanar straight segments is the mutual contribution
from non-adjacent straight segments, see figure (\ref{Fig_NonAdjacent}).
\begin{eqnarray*}
L\left(m,m^{\prime},n,n^{\prime},\alpha\right) & = & \frac{\mu_{0}}{2\pi}\left\{ A\left(m^{\prime},n^{\prime},n,\alpha\right)+A\left(n^{\prime},m^{\prime},m,\alpha\right)+A\left(m,n,n^{\prime},\alpha\right)+A\left(n,m,m^{\prime},\alpha\right)\right\} ,\\
A\left(w,u,v,\alpha\right) & = & \left[w\operatorname{asinh}\left(\frac{u+w\cos\alpha}{w\sin\alpha}\right)-w\operatorname{asinh}\left(\frac{v+w\cos\alpha}{w\sin\alpha}\right)\right]\cos\alpha.
\end{eqnarray*}
This leads to an unwieldy expression already for a hexagon, but the
calculation of the self inductance of such loops is a matter of algebra
and geometry (not a new result).

\bigskip{}

\end{document}